\documentclass[acmsmall]{acmart}

\usepackage[linesnumbered,ruled,vlined]{algorithm2e}
\usepackage{multirow}
\usepackage{wrapfig}

\definecolor{codegreen}{rgb}{0,0.4,0}
\definecolor{codegray}{rgb}{0.5,0.5,0.5}
\definecolor{codepurple}{rgb}{0.58,0,0.82}
\definecolor{backcolour}{rgb}{0.95,0.95,0.92}

\usepackage{listings}

\lstnewenvironment{code}
  {\lstset{language=C,
    commentstyle=\color{codegreen},
    keywordstyle=\color{blue}\bfseries,
    numberstyle=\tiny\color{codegray},
    stringstyle=\color{codepurple},
    aboveskip=0.5\medskipamount,
    belowskip=0.5\medskipamount,
    linewidth=\textwidth,
    xleftmargin=0.1\textwidth,
    xrightmargin=0.1\textwidth,
    breakatwhitespace=false,
    breaklines=true,
    keepspaces=true,
    showspaces=false,
    showstringspaces=false,
    showtabs=false,
    tabsize=2,
    basicstyle=\ttfamily\scriptsize,
    morekeywords={sunindextype,
                  realtype,
                  booleantype,
                  N_Vector,
                  N_Vector_Ops,
                  N_VectorContent_Serial,
                  SUNMatrix,
                  SUNLinearSolver,
                  SUNLinearSolver_Ops,
                  SUNLinearSolver_Type,
                  SUNNonlinearSolver,
                  SUNNonlinearSolver_Type,
                  SUNNonlinSolSysFn,
                  SUNNonlinSolLSetupFn,
                  SUNNonlinSolLSolveFn,
                  SUNNonlinSolConvTestFn}
    }%
  }
  {}

\lstnewenvironment{codetablemath}
  {\vspace{-0.8\baselineskip}
   \lstset{language=C,
    aboveskip=0pt,
    belowskip=0pt,
    basicstyle=\ttfamily\scriptsize,
    breaklines=true,
    morekeywords={sunindextype,
                  realtype,
                  booleantype,
                  N_Vector,
                  N_Vector_Ops,
                  N_VectorContent_Serial,
                  SUNMatrix,
                  SUNLinearSolver,
                  SUNLinearSolver_Type,
                  SUNNonlinearSolver,
                  SUNNonlinearSolver_Type,
                  SUNNonlinSolSysFn,
                  SUNNonlinSolLSetupFn,
                  SUNNonlinSolLSolveFn,
                  SUNNonlinSolConvTestFn}
    }%
  }
  {\vspace{-0.8\baselineskip}}

\lstnewenvironment{codetable}
  {\vspace{-0.6\baselineskip}
   \lstset{language=C,
    aboveskip=0pt,
    belowskip=0pt,
    basicstyle=\ttfamily\scriptsize,
    breaklines=true,
    morekeywords={sunindextype,
                  realtype,
                  booleantype,
                  N_Vector,
                  N_Vector_Ops,
                  N_VectorContent_Serial,
                  SUNMatrix,
                  SUNLinearSolver,
                  SUNLinearSolver_Type,
                  SUNNonlinearSolver,
                  SUNNonlinearSolver_Type,
                  SUNNonlinSolSysFn,
                  SUNNonlinSolLSetupFn,
                  SUNNonlinSolLSolveFn,
                  SUNNonlinSolConvTestFn}
    }%
  }
  {\vspace{-0.6\baselineskip}}

\newcommand{\editmade}[1]{\textcolor{black}{#1}}

\newcommand*\wraptt[1]{{\small\texttt{\expandafter\ttbreak\detokenize{#1}\relax}}}
\newcommand*\ttbreak[1]{\ifx\relax#1\else
  \expandafter\ifx\string_#1\string_\allowbreak\else#1\fi
  \expandafter\ttbreak\fi}

\AtBeginDocument{%
  \providecommand\BibTeX{{%
    \normalfont B\kern-0.5em{\scshape i\kern-0.25em b}\kern-0.8em\TeX}}}

\setcopyright{usgovmixed}
\copyrightyear{2021}
\acmYear{2021}
\acmDOI{00.0000/0000000.0000000}

\acmJournal{TOMS}
\acmVolume{0}
\acmNumber{0}
\acmArticle{0}
\acmMonth{0}

\begin{document}

\title[Enabling New Flexibility in SUNDIALS]{Enabling New Flexibility in the SUNDIALS Suite of Nonlinear and Differential/Algebraic Equation Solvers}

\author{David J. Gardner}
\orcid{0000-0002-7993-8282}
\affiliation{%
    \institution{Lawrence Livermore National Laboratory}
    \department{Center for Applied Scientific Computing}
    \city{Livermore}
    \state{California}}
\email{gardner48@llnl.gov}

\author{Daniel R. Reynolds}
\orcid{0000-0002-0911-7841}
\affiliation{%
    \institution{Southern Methodist University}
    \department{Department of Mathematics}
    \city{Dallas}
    \state{Texas}}
\email{reynolds@smu.edu}

\author{Carol S. Woodward}
\orcid{0000-0002-6502-8659}
\affiliation{%
    \institution{Lawrence Livermore National Laboratory}
    \department{Center for Applied Scientific Computing}
    \city{Livermore}
    \state{California}}
\email{woodward6@llnl.gov}

\author{Cody J. Balos}
\orcid{0000-0001-9138-0720}
\affiliation{%
    \institution{Lawrence Livermore National Laboratory}
    \department{Center for Applied Scientific Computing}
    \city{Livermore}
    \state{California}}
\email{balos1@llnl.gov}

\renewcommand{\shortauthors}{D.~J.~Gardner, D.~R.~Reynolds, C.~S.~Woodward, and C.~J.~Balos}

\authorsaddresses{%
This research was supported by the \grantsponsor{ECP}{Exascale Computing Project (ECP)}{https://exascaleproject.org/}, Project Number: \grantnum{ECP}{17-SC-20-SC}, a collaborative effort of the U.S. Department of Energy Office of Science and the National Nuclear Security Administration.
Support for this work was also provided through the \grantsponsor{SciDAC}{Scientific Discovery through Advanced Computing (SciDAC) project}{https://www.scidac.gov/} ``Frameworks, Algorithms and Scalable Technologies for Mathematics (FASTMath),'' funded by the U.S. Department of Energy Office of Advanced Scientific Computing Research and National Nuclear Security Administration.
This work was performed under the auspices of the U.S. Department of Energy by Lawrence Livermore National Laboratory under contract DE-AC52-07NA27344. Lawrence Livermore National Security, LLC.  LLNL-JRNL-816631.
Author's addresses:
D.~J.~Gardner, C.~S.~Woodward, and C.~J.~Balos, Center for Applied Scientific Computing, Lawrence Livermore National Laboratory, 7000 East Ave, L-561, Livermore, CA 94550; emails: \{gardner48, woodward6, balos1\}@llnl.gov;
D.~R.~Reynolds, Department of Mathematics, Southern Methodist University, P.O. Box 750156, Dallas, TX 75275; email: reynolds@smu.edu.}

\begin{abstract}
  In recent years, the SUite of Nonlinear and DIfferential/ALgebraic equation Solvers (SUNDIALS) has been redesigned to better enable the use of application-specific and third-party algebraic solvers and data structures. Throughout this work, we have adhered to specific guiding principles that minimized the impact to current users while providing maximum flexibility for later evolution of solvers and data structures. The redesign was done through \editmade{the addition of new linear and nonlinear solvers classes}, enhancements to the vector class, and the creation of modern Fortran \editmade{interfaces.} The vast majority of this work has been performed ``behind-the-scenes,'' with minimal changes to the user interface and no reduction in solver capabilities or performance. \editmade{These changes allow SUNDIALS users to more easily utilize external solver libraries and create highly customized solvers, enabling greater flexibility on extreme-scale, heterogeneous computational architectures.}
\end{abstract}



\maketitle

\section{Introduction}
\label{sec:intro}
The SUite of Nonlinear and DIfferential/ALgebraic equation Solvers (SUNDIALS) \cite{hindmarsh2005sundials} is a collection of software packages designed to solve time-dependent and nonlinear equations on large-scale, high performance computing \editmade{(HPC)} systems. The packages have evolved from a multi-decade history of highly efficient solution packages for ordinary differential equations (ODEs) \cite{hindmarsh1980lsode,hindmarsh1983odepack,brown1989vode}, differential-algebraic equations (DAEs) \cite{petzold1982description}, nonlinear systems \cite{brown1990hybrid,brown1994using}, and sensitivity analysis enabled integrators \cite{li2000software,serban2005cvodes}. The SUNDIALS codes take advantage of the many innovations this history and the methods community have delivered in terms of adaptive time step and adaptive order integrators for ODEs and DAEs \cite{byrne1975polyalgorithm,jackson1980alternative, brown1998consistent,hairer1993solving,hairer1996solving,kennedy2003additive}, the combination of Krylov and Newton methods for efficient nonlinear solvers \cite{brown1986matrix,brown1989reduced,pernice1998nitsol}, methods and software for sensitivity analysis \cite{maly1996numerical,feehery1997efficient,li2000software,cao2003adjoint}, \editmade{and effective solver} controls for time integrators and nonlinear solvers \cite{hindmarsh1988krysi,hindmarsh1992detecting,byrne1987stiff,hindmarsh1995algorithmsP1,hindmarsh1995algorithmsP2}. \editmade{Over the last two decades the SUNDIALS codes have been used in a number of applications on small to large-scale computing systems \cite{gardner2015implicit, mcnenly2015faster, gardner2018implicit, vogl2019evaluation, kollet2010proof, dorr2010numerical, dudson2009bout++, zhang2020amrex, carnevale2006neuron, goodwin2018cantera} as well as in other mathematical libraries and frameworks \cite{rackauckas2017differentialequations, andersson2019casadi, anderson2020mfem, malengier2018odes, arndt2020deal, schmidt2007sbaddon, wolfram2020ndsolve, zhang2020amrex}, garnering more than 90,000 downloads in 2020.}

\editmade{During this time, the science simulated as well as the software needed to carry out those simulations have grown in complexity as HPC architectures have transitioned from simplistic homogeneous systems to heterogeneous systems (e.g., CPU+GPU). With the trend toward heterogeneity, the object-oriented design of the SUNDIALS vector structure \cite{hindmarsh2005sundials} has enabled incorporating new HPC programming models \cite{karlin2019preparation,reynolds2019sundials}. However, creating interfaces to utilize new algebraic solver packages targeting large-scale heterogeneous systems would require extensive redundant code with limited extensibility. This paper describes recent enhancements following an object-oriented design to ensure adaptability to new machines and programming environments, allow users to construct architecture-aware and problem-specific data structures and solvers, ease interoperability with other libraries, and provide greater flexibility to users and for future extensions. The design principles that guided this work have been: assume as little about data layouts as possible; define interfaces that allow users to provide their own data structures and solvers with little new code; ensure SUNDIALS packages can evolve and change their offerings under these interfaces; produce highly sustainable code; and minimize the impact to current users.}

The rest of this paper is organized as follows. The next section gives a \editmade{brief overview} of the SUNDIALS packages and is followed by a discussion \editmade{of the core SUNDIALS classes.} Section \ref{sec:vectors} describes changes to the vector class and \editmade{new vector} implementations. Sections \ref{sec:linear} and \ref{sec:nonlinear} discuss new classes for linear and nonlinear solvers, respectively \editmade{with accompanying demonstration programs.} Section \ref{sec:f2003} overviews the modern\editmade{, sustainable Fortran 2003 interfaces to SUNDIALS.} Lastly, Section \ref{sec:conclusions} gives a summary and next steps for SUNDIALS.

Throughout the text, the presentation of new features and enhancements is generally focused at the level of interfaces and functionality while many of the implementation-specific details of the contributions are excluded. However, references to detailed descriptions in the SUNDIALS user guides are provided for the interested reader. As the user guide for each SUNDIALS package includes shared text regarding the overall SUNDIALS infrastructure, these citations refer to the current CVODE user guide \cite{cvodeDocumentation}. \editmade{For details on updating existing codes to the latest SUNDIALS version, see the ``Recent Changes'' section of the package's user guide and the examples included with each release.} The SUNDIALS source code\editmade{, example programs,} and user guides can be obtained from the SUNDIALS web page \cite{sundials-web} or GitHub repository \cite{sundials-github}.

\section{\protect\editmade{SUNDIALS Overview}}
\label{sec:overview}

\editmade{This section provides a brief overview of the numerical methods implemented in the six SUNDIALS packages: CVODE, CVODES, IDA, IDAS, ARKODE, and KINSOL. Additional details on each package, with the exception of ARKODE, are available in \citeN{hindmarsh2005sundials}. As the native vector and matrix derived classes are written for problems in real $N$-space, the following overview considers problems posed in $\mathbb{R}^N$. Problems posed in other spaces, e.g., $\mathbb{C}^N$, can be solved using user-defined derived classes.}

\subsection{\editmade{CVODE and CVODES}}

CVODE targets stiff and nonstiff initial value problems (IVPs) for ODEs in the explicit \editmade{form}
\begin{equation*}
    \dot{y} = f(t,y), \quad y(t_0) = y_0,
\end{equation*}
where $t$ is the independent variable (e.g., time), the dependent variable is $y \in \mathbb{R}^N$, $\dot{y}$ denotes $\mathrm dy/\mathrm dt$, and $f: \mathbb{R} \times \mathbb{R}^{N} \to \mathbb{R}^N$. CVODE uses variable order and variable step size implicit linear multistep methods in the fixed-leading-coefficient form \cite{byrne1975polyalgorithm, jackson1980alternative} given by
\begin{equation*}
 \sum_{i = 0}^{K_1} \alpha_{n,i} y_{n-i} +
     h_n \sum_{i = 0}^{K_2} \beta_{n,i} \editmade{f(t_{n-i}, y_{n-i})} = 0,
\end{equation*}
where $y_n$ is a computed approximation to $y(t_n)$, $h_n = t_n - t_{n-1}$ is the time step size, the values $K_1$ and $K_2$ are determined by the method type and order of accuracy, and the coefficients $\alpha_{n,i}$ and $\beta_{n,i}$ are uniquely determined by the method type and order as well as the recent step size history. \editmade{Adams-Moulton methods of order 1 to 12 are provided for non-stiff problems and Backward Differentiation Formulas (BDF) of order 1 to 5 for stiff cases.} Additionally, CVODE supports projection methods \cite{eich1993convergence, shampine1999conservation} for integrating ODE systems with constraints i.e., \editmade{where} the solution must satisfy $g(t,y) = 0$.

CVODES \cite{serban2005cvodes} is an extension of CVODE for conducting forward or adjoint sensitivity analysis of ODE IVPs  with respect to the parameters $p \in \mathbb{R}^{N_p}$, i.e., problems of the form,
\begin{equation*}
    \dot{y} = f(t,y,p), \quad y(t_0) = y_0(p).
\end{equation*}

\subsection{\editmade{IDA and IDAS}}

IDA targets \editmade{the more general case of DAE and implicit ODE IVPs in the form}
\begin{equation*}
    F(t,y,\dot{y}) = 0, \quad y(t_0) = y_0, \quad \dot{y}(t_0) = \dot{y}_0\editmade{.}
\end{equation*}
\editmade{The state is evolved using variable order and variable step size BDF methods in fixed-leading-coefficient form \cite{brenan1995numerical} of order $q = 1, \dots, 5$ given by,}
\begin{equation*}
  \sum_{i=0}^q \alpha_{n,i}y_{n-i} = h_n \dot{y}_n.
\end{equation*}
As with CVODES and CVODE, IDAS is an extension to IDA with forward and adjoint sensitivity analysis capabilities for DAE \editmade{and implicit ODE} IVPs of the form
\begin{equation*}
    F(t,y,\dot{y},p) = 0, \quad y(t_0) = y_0(p), \quad \dot{y}(t_0) = \dot{y}_0(p).
\end{equation*}

\subsection{\editmade{ARKODE}}

\editmade{The ARKODE package targets ODE IVPs in the linearly-implicit form}
\begin{equation*}
    M(t)\, \dot{y} = f(t,y), \quad \quad y(t_0) = y_0,
\end{equation*}
where for any $t$, $M(t):\mathbb{R}^{N} \to \mathbb{R}^{N}$ is a nonsingular linear \editmade{operator. In many uses cases $M(t)$ is the identity matrix while in finite element computations it is an $N \times N$ mass matrix (or a function for its action on a vector).} The right-hand side function may be additively partitioned as $f(t,y) = f_1(t,y) + f_2(t,y)$.

\editmade{For} stiff, nonstiff, and mixed stiff/nonstiff problems, where $f_1 = f^E$ may contain the nonstiff (explicit) components and $f_2 = f^I$ the stiff (implicit) components of the system respectively, ARKODE provides explicit, implicit, and implicit-explicit (IMEX) additive Runge-Kutta methods \cite{ascher1995implicit,ascher1997implicit,kennedy2003additive,kennedy2019higher} with the general form
\begin{equation*}
\begin{split}
    z_i &= y_{n-1} + h_n \sum_{j=1}^{i-1} A^E_{i,j} \hat{f}^E(t_{n-1} + c^E_j h_n, z_j)
              + h_n \sum_{j=1}^{i} A^I_{i,j} \hat{f}^I(t_{n-1} + c^I_j h_n, z_j),
\quad i=1,\ldots,s, \\
    y_n &= y_{n-1} + h_n \sum_{i=1}^{s} \left(b^E_i \hat{f}^E(t_{n-1} + c^E_j h_n, z_i)
              + b^I_i \hat{f}^I(t_{n-1} + c^I_j h_n, z_i)\right), \\
    \tilde{y}_n &= y_{n-1} + h_n \sum_{i=1}^{s} \left(
               \tilde{b}^E_i \hat{f}^E(t_{n-1} + c^E_j h_n, z_i) +
               \tilde{b}^I_i \hat{f}^I(t_{n-1} + c^I_j h_n, z_i)\right),
\end{split}
\end{equation*}
where $\hat{f}^E(t,y) = M(t)^{-1}f^E(t,y)$ and $\hat{f}^I(t,y) = M(t)^{-1}f^I(t,y)$ convert the IVP from linearly implicit to explicit form, $z_i$ are internal method stages, and $\tilde{y}_n$ is an embedded solution for temporal error estimation.
The number of stages \editmade{$s$} and the coefficients $A^* \in \mathbb{R}^{s\times s}$, $b^* \in \mathbb{R}^{s}$, $c^* \in \mathbb{R}^{s}$, and $\tilde{b}^* \in \mathbb{R}^{s}$ are given by the explicit and implicit Butcher tables that define the method \editmade{and its embedding}.

\editmade{For} multirate problems, the right-hand side may be split into ``slow'' components, $f_1 = f^S$, advanced with a large step $H_n$ and ``fast'' components, $f_2 = f^F$, advanced with time step, $h_n \ll H_n$.  For such problems, ARKODE provides multirate infinitesimal step \cite{schlegel2009multirate,schlegel2012numerical,schlegel2012implementation} and multirate infinitesimal GARK \cite{sandu2019class} \editmade{methods. These methods} utilize an $s$-stage Runge-Kutta method where the stages $z_i$ are computed by solving (with a possibly different method) the auxiliary ODE IVP
\begin{equation*}
\begin{aligned}
    \dot{v}(\theta) &= \Delta c_i f^F(T_{i-1}+\Delta c_i\theta, v(\theta)) + r(\theta), \quad \theta \in [0, H_n],\\
    v(0) &= z_{i-1},\notag
\end{aligned}
\end{equation*}
where $T_i = t_n+c_i H_n$, $\Delta c_i=c_i-c_{i-1}$, and $r(\theta)$ is a ``forcing function'' constructed from evaluations of $f^S(T_j,z_j)$, $j=1,\ldots,i$, that propagates information from the slow time scale. The fast IVP solution is used for the new internal stage i.e., $z_i = v(H_n)$.

\subsection{\editmade{KINSOL}}

Finally the KINSOL package solves nonlinear systems in root-finding or fixed-point form,
\begin{equation}
\label{eq:rootfind_vs_fixedpoint}
    F(u) = 0 \quad \text{and} \quad G(u) = u
\end{equation}
respectively, where $u \in \mathbb{R}^N$, and the functions $F$ and $G$ map $\mathbb{R}^N \to \mathbb{R}^N$.
\editmade{For} the root-finding case, KINSOL provides inexact and modified Newton methods \cite{dembo1982inexact,ortega2000iterative} which may optionally utilize a line search \editmade{strategy\cite{dennis1996numerical}. For} general fixed-point systems and systems with the special form $F(u) = Lu - \mathcal{N}(u) \Rightarrow G(u) = u - L^{-1} F(u)$, where $L$ is a nonsingular linear operator (an $N \times N$ matrix or a function for its action on a vector) and $\mathcal{N}$ is (in general) nonlinear, KINSOL implements fixed-point and Picard iterations \editmade{$u_{n+1} = G(u_n)$.}
The convergence of the iteration may be significantly accelerated by optionally applying Anderson's method \cite{anderson1965iterative,walker2011anderson}.

\subsection{\editmade{Integrator Nonlinear Systems}}
\editmade{At each time step the implicit integration methods in CVODE(S), ARKODE, and IDA(S) must solve one or more nonlinear systems i.e.,}
\begin{align*}
    \text{CVODE:}  \quad & y_{n} - h_n \beta_{n,0} f(t_n,y_n) - a_n = 0,\\
    \text{ARKODE:} \quad & z_{i} - h_n A_{i,i}^I \hat{f}^{I}(t_{n-1} + c^I_j h_n, z_i) - a_i = 0, \text{ and}\\
    \text{IDA:}    \quad & F\left(t_n, y_{n}, h_n^{-1} {\textstyle\sum}_{i=0}^q \alpha_{n,i} y_{n-i}\right) = 0.
\end{align*}
\editmade{where $a_*$ is comprised of known data from prior time steps or stages. In adjoint sensitivity computations the same nonlinear systems are solved in the backward problem while forward sensitivity computations require solving an expanded nonlinear system that includes sensitivity equations. In all cases, these nonlinear systems can be viewed as generic root-finding or generic fixed-point problems and are solved using many of the same methods employed by KINSOL.}

\editmade{Note that in the forward sensitivity case CVODES and IDAS utilize the strategy discussed in \citeN{maly1996numerical} where a block diagonal approximation of the combined system Jacobian is used when solving the larger nonlinear system with Newton's method. This decoupling enables reusing the ODE/DAE system Jacobian without additional matrix factorizations or preconditioner setups. CVODES and IDAS leverage this structure to utilize the same nonlinear solvers and linear solver interfaces employed by the non-sensitivity-enabled versions of the integrators.}

\section{SUNDIALS Structure}
\label{sec:structure}

\begin{figure}[htb!]
    \centering
    \includegraphics[width=0.90\textwidth]{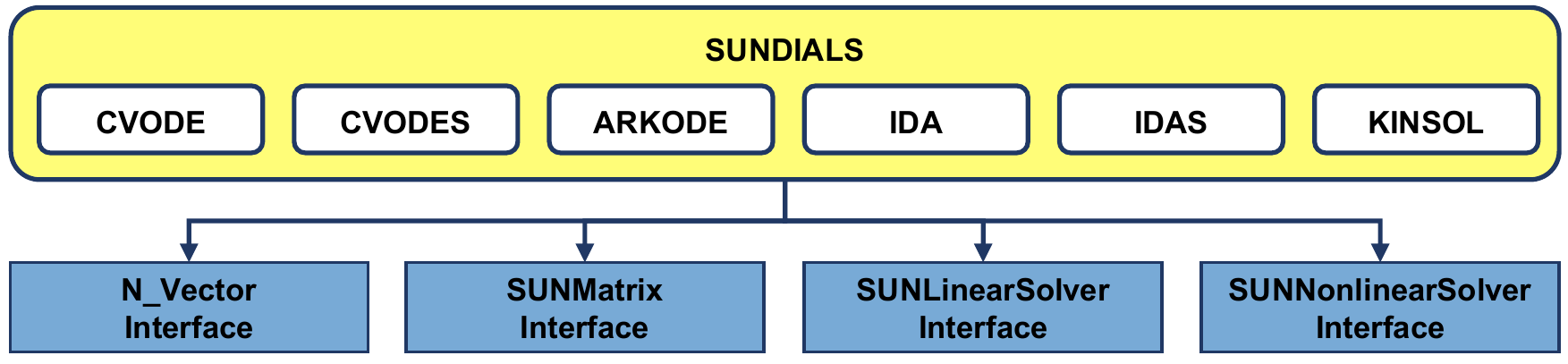}
    \caption{\editmade{An overview of the SUNDIALS packages (CVODE, CVODES, IDA, IDAS, ARKODE, and KINSOL) and the four base classes (N\_Vector, SUNMatrix, SUNLinearSolver, and SUNNonlinearSolver).}}
    \label{fig:sunorg}
\end{figure}

\editmade{To} implement the above methods for integrating ODEs and DAEs or solving nonlinear systems, one must be able to perform the following actions:
\begin{enumerate}
    \item Compute vector operations e.g., add vectors, compute norms, etc.
    \item Solve the nonlinear systems that arise in implicit integration methods.
    \item Solve the linear systems that arise within nonlinear solvers or linearly-implicit ODEs (i.e., mass matrix linear systems).
    \item Compute matrix operations e.g., to explicitly construct the linear systems when necessary or to compute matrix-vector products.
\end{enumerate}
These \editmade{shared} requirements \editmade{across the packages} lead directly to the \editmade{SUNDIALS N\_Vector, SUNMatrix, SUNLinearSolver, and SUNNonlinearSolver classes shown in Figure \ref{fig:sunorg} that define abstract interfaces for performing vector and matrix operations as well as solving linear and nonlinear systems, respectively.}
\editmade{All of the packages and the solver interfaces depend on the vector interface; however, connections between the other classes depend on the particular derived class and use case e.g., matrix-free linear solvers may accept a \texttt{NULL} matrix object.}

\editmade{This object-oriented design encapsulates implementation-specific details behind a set of well defined abstract interfaces for performing mathematical operations. This encapsulation enables writing the SUNDIALS packages purely in terms of generic operations on objects, making them agnostic to the user's particular choice of data structures/layout, method of parallelization, algebraic solvers, etc. Moreover, the design provides flexibility, eases the addition of new features, and maximizes the reuse of shared code.
With the new class interfaces discussed in this paper, SUNDIALS packages more seamlessly work with other libraries that provide complementary capabilities for scientific simulations. For example, SUNDIALS can interoperate with discretization and solver libraries as long as class implementations are provided for interfacing to other library’s data structures and, if necessary,  solvers.}

\subsection{\editmade{Previous SUNDIALS structure}}

\editmade{While the vector class has long been a feature of SUNDIALS, this object-oriented approach was not originally applied to matrices or algebraic solvers.
Earlier versions of SUNDIALS did clearly separate the linear solvers and matrices from the integrators and nonlinear solvers.  However, these interfaces suffered from three primary challenges. First, each SUNDIALS package defined its own unique interface to each linear solver implementation, leading to significant code duplication and an unnecessary maintenance burden. Second, since each package had a separate linear solver API, it was not possible to provide a custom or third-party linear solver that could be immediately utilized by all the SUNDIALS packages, e.g., a custom linear solver written for CVODE could not be used by ARKODE without writing additional interface code. Lastly, providing the interface functions and performing some computations required directly accessing private data within the packages, which hindered the ability of the packages to evolve without breaking user code.}

\editmade{Unlike the linear solvers, earlier versions of SUNDIALS did not separate the nonlinear solvers from the integrators. While the integrator-specific nonlinear systems could be written in a generic form, the nonlinear solvers were embedded within the integrators. As such, each package implemented its own nonlinear solvers and did not provide a means for interfacing with external or user-defined nonlinear solvers leading to duplicated code and inconsistent solver options between the packages.
Moreover, the lack of a shared nonlinear solver interface across the integration packages limited their flexibility and increased the software maintenance burden.}

\editmade{To address the challenges presented by the old linear solver interfaces and the absence of a nonlinear solver interface, we completely redesigned the handling of matrices, linear solvers, and nonlinear solvers throughout SUNDIALS. Our specific goals in this redesign were to:
\begin{itemize}
\item Collapse the separate package-specific linear solver interfaces into a single generic interface, thereby removing redundant code.
\item Unify the package-specific nonlinear solver implementations into generic implementations shared across the integrators.
\item Maintain the ability to provide integrator-specific controls over linear and nonlinear solvers for performance optimization.
\item Provide streamlined interfaces with clearly defined base classes (and examples) so users can easily leverage application-specific or external algebraic solvers.
\end{itemize}}

\subsection{\editmade{SUNDIALS Class Structure}}
\label{sec:sundials-class-structure}

\editmade{All of the SUNDIALS base classes follow the same fundamental design and are defined as a C structure containing a \texttt{void} pointer to the derived class member data and a virtual method table (VMT) implemented as a C structure of function pointers. For example, the new SUNLinearSolver class is defined in \href{https://github.com/LLNL/sundials/blob/master/include/sundials/sundials_linearsolver.h}{\wraptt{include/sundials/sundials_linearsolver.h}} as follows}
\begin{center}
\begin{code}
typedef struct _generic_SUNLinearSolver *SUNLinearSolver;
typedef struct _generic_SUNLinearSolver_Ops *SUNLinearSolver_Ops;

struct _generic_SUNLinearSolver {
  void *content;            // pointer to member data
  SUNLinearSolver_Ops ops;  // virtual method table
};
\end{code}
\end{center}
\editmade{Note the base class VMT includes pure virtual functions for which a derived class must provide an implementation, virtual functions the derived class may override, and optional functions that, if non-\texttt{NULL}, the SUNDIALS packages will utilize to improve performance. Thus, the base class pure virtual functions define the required minimum set of operations for derived classes.}

\editmade{To aid users in creating derived class implementations, constructors are provided to allocate and initialize the base class structure. After creating the base class, the derived class is responsible for any member data allocations, setting the base class member data pointer, and setting the function pointers in the VMT. In the linear solver case, the constructor for a derived class would proceed approximately as follows}
\begin{center}
\begin{code}
SUNLinearSolver MySUNLinearSolver(...) {
  SUNLinearSolver LS = SUNLinSolNewEmpty(); // create the base class
  LS->content        = my_member_data;      // set member data pointer
  LS->ops->solve     = my_solve_function;   // set method pointers
  // ...
  return LS;
}
\end{code}
\end{center}
\editmade{Additionally, base class methods are provided to copy the VMT when cloning objects thus ensuring the method pointers are set correctly. The base constructors and utility methods methods enable the introduction of new specialized virtual and optional methods aimed at increasing the capability and/or performance of the class with minimal impact on current users. Furthermore, when combined with the new Fortran interfaces discussed in Section \ref{sec:f2003}, these methods enable the creation of custom SUNDIALS class implementations from Fortran.}

\section{Enhanced Vector Structures}
\label{sec:vectors}

\editmade{In} the years since SUNDIALS was first released, \editmade{the vector} encapsulation has served an invaluable role in allowing existing codes to rapidly utilize SUNDIALS integrators and solvers, since users needed only to provide an implementation of the N\_Vector class to wrap \editmade{and operate on} their application-specific data \editmade{structures. To} adapt to a wider array of parallel computing strategies than only pure MPI, SUNDIALS has been progressively updated to provide a wider set of N\_Vector implementations than the original serial and MPI-parallel versions, \editmade{and now} includes vectors based on OpenMP, Pthreads, CUDA, \editmade{HIP}, \editmade{SYCL}, and RAJA for various forms of on-node parallelism. Additionally, implementations that wrap vector objects from PETSc, \textit{hypre}, and Trilinos have been added to streamline interfacing with other mathematical libraries. \editmade{However, over that time the required set of vector operations remained static, and there was not a native vector implementation that supported collecting independent vectors into a single object.}

As modern computing hardware increasingly includes mixtures of different processing elements (multi-core CPU, GPU, FPGA, etc.), and since floating-point computing speeds have far outpaced memory access speeds, an update to the underlying N\_Vector structure was needed. \editmade{We thus extended the vector class in two distinct ways:
\begin{enumerate}
    \item Expanded the set of vector methods to include operations that reduce the number of kernel launches, increase arithmetic intensity, and/or minimize parallel communication.
    \item Added a ``many-vector'' implementation of the vector base class that enables combining different independent vector instances, each using potentially different parallelization approaches targeting distinct pieces of hardware, into a single vector.
\end{enumerate}}

\subsection{New Vector Operations}
\label{subsec:VectorOps}

Throughout a simulation using SUNDIALS, there exist numerous locations that use the same sets of repeated vector operations.  Each of these operations over a vector of length $N$ requires $\mathcal{O}(N)$ floating-point operations and memory accesses. Additionally, for accelerator-based implementations, each vector operation requires a separate kernel launch, and for MPI-based parallel vectors each dot product or norm-like operation requires one \texttt{MPI\_Allreduce} call.  While the total floating-point work of these sequences of vector operations has remained unchanged, we have now reduced these auxiliary costs (memory accesses, kernel launches, and MPI reductions) by creating a set of ``fused'' operations that each combine a specific sequence of operations into a single routine. These fused operations thus provide the potential for cost savings by increasing arithmetic intensity and reducing overhead.

\begin{table}[htb]
\caption{\editmade{Subset of optional fused and vector array operations.}}
\label{t:fused_vec_ops}
\lstset{linewidth=0.70\textwidth}
\small
\begin{tabular}{p{0.70\textwidth}p{0.20\textwidth}}
\toprule
Function & Description \\
\midrule
\begin{codetablemath}
int N_VLinearCombination(int n, realtype* c, N_Vector* X,
                         N_Vector z)
\end{codetablemath}
&
$\begin{array}{c}
    z = \displaystyle\sum_{i=0}^{n-1} c_i\, X_i
\end{array}$ \\
\midrule
\begin{codetablemath}
int N_VScaleAddMulti(int n, realtype* a, N_Vector x, N_Vector* Y,
                     N_Vector* Z)
\end{codetablemath}
& $\begin{array}{c}
    Z_i = a_i\, x + Y_i \\
    \text{for } i = 0, \ldots, n-1
\end{array}$ \\
\midrule
\begin{codetablemath}
int N_VDotProdMulti(int n, N_Vector x, N_Vector* Y, realtype* d)
\end{codetablemath}
&
$\begin{array}{c}
    d_i = x \cdot Y_i \\
    \text{for } i = 0, \ldots, n-1
\end{array}$ \\
\midrule
\begin{codetablemath}
int N_VLinearSumVectorArray(int n, realtype a, N_Vector* X,
                            realtype b, N_Vector* Y, N_Vector* Z)
\end{codetablemath}
&
$\begin{array}{c}
     Z_i = a\, X_i + b\, Y_i \\
     \text{for } i = 0, \ldots, n-1
\end{array}$ \\
\bottomrule
\end{tabular}
\end{table}

We group these new operations into two categories. The first set\editmade{, shown in the first three rows of Table \ref{t:fused_vec_ops},} fuses standard vector operations into a single operation. \editmade{The \texttt{N\_VLinearCombination} and \texttt{N\_VScaleAddMulti} operations replace multiple calls to \texttt{N\_VLinearSum}, which computes $z = a x + b y$, with a single call. Similarly, \texttt{N\_VDotProdMulti} combines the reductions from several \texttt{N\_VDotProd} calls, which compute $x \cdot y$, into a single reduction. The second category is composed of operations on collections of vectors that arise in sensitivity analysis simulations. An example is shown in the last row of Table \ref{t:fused_vec_ops} where the linear sum of multiple vectors are computed in one operation.}

To minimize the requirements when supplying a custom N\_Vector implementation, \editmade{all of these new operations are \textit{optional} as the base class implementation calls the pre-existing sequence of \textit{required} operations to achieve the same result.
Thus, the operations can be enabled or disabled at runtime by setting the VMT function pointer to the implementation defined function or to \texttt{NULL}, thereby allowing the user to implement only the operations that are most impactful for their use case.}
More details on the fused operations and the full set of vector array operations that have been added to the class can be found in \cite[Sections 6.1.2 and 6.1.3]{cvodeDocumentation}.

\editmade{In standalone CPU-based vector tests, we have found that fused streaming operations are beneficial in cases that work on many vectors, e.g., with high-order integrators with more than approximately 1,000 unknowns per MPI task. Fused reduction operations arise in orthogonalization methods in iterative methods and are beneficial for any number of vectors and with less than approximately 130K unknowns per MPI task. We note these numbers are machine dependent and will certainly change as architectures advance.}

\subsection{Many-vector}
\label{subsec:ManyVector}

To enable SUNDIALS packages to leverage heterogeneous computational architectures, increase the ability for asynchronous computation, and enable greater flexibility for multiphysics applications, we have created two ``many-vector'' \editmade{implementations of the N\_Vector class. These vectors are a software layer to enable operating on a collection of distinct vector instances as if they were a single cohesive vector. The many-vector implementations} do not directly manipulate vector data\editmade{; all computations are carried out by the underlying vector instances.} As such, the many-vectors are designed to provide users with more complete control over where data is laid out on parallel and heterogeneous architectures, and, in turn, over which resources perform calculations on this data. At a high-level, the design is based on the following objectives:
\begin{itemize}
\item Provide a clean mechanism for users to partition their simulation data among disparate computational resources (different nodes, different compute architectures, different types of memory, etc.), enabling ``hybrid'' computations on heterogeneous architectures.
\item Facilitate the use of a separate MPI communicator for the collection than is used for any of its subvectors, allowing the higher-level communicator to act as an MPI ``intercommunicator'' enabling coupled simulations between codes that utilize disjoint MPI communicator groups. From a multiphysics perspective, these disjoint groups can correspond to fully-functioning MPI-based simulations, each using its own standard MPI intracommunicator -- the intercommunicator enables messages to be sent between these disjoint simulations.
\item Allow computationally efficient user-supplied routines to perform operations on the relevant subvectors.
\item Enable the creation of new SUNDIALS class implementations that leverage the data partitioning (e.g., partitioned integration methods, block (non)linear solvers, etc.) to improve performance for particular application types.
\item Ensure seamless functionality with any existing SUNDIALS package or class.
\end{itemize}

We envision two non-mutually-exclusive scenarios where the many-vector capability will benefit users \editmade{(though other use cases are possible)}: partitioning data between different computing \editmade{resources and} combining distinct MPI intracommunicators together into a multiphysics simulation.

\begin{figure}[b!]
    \centering
    \includegraphics[width=0.75\textwidth]{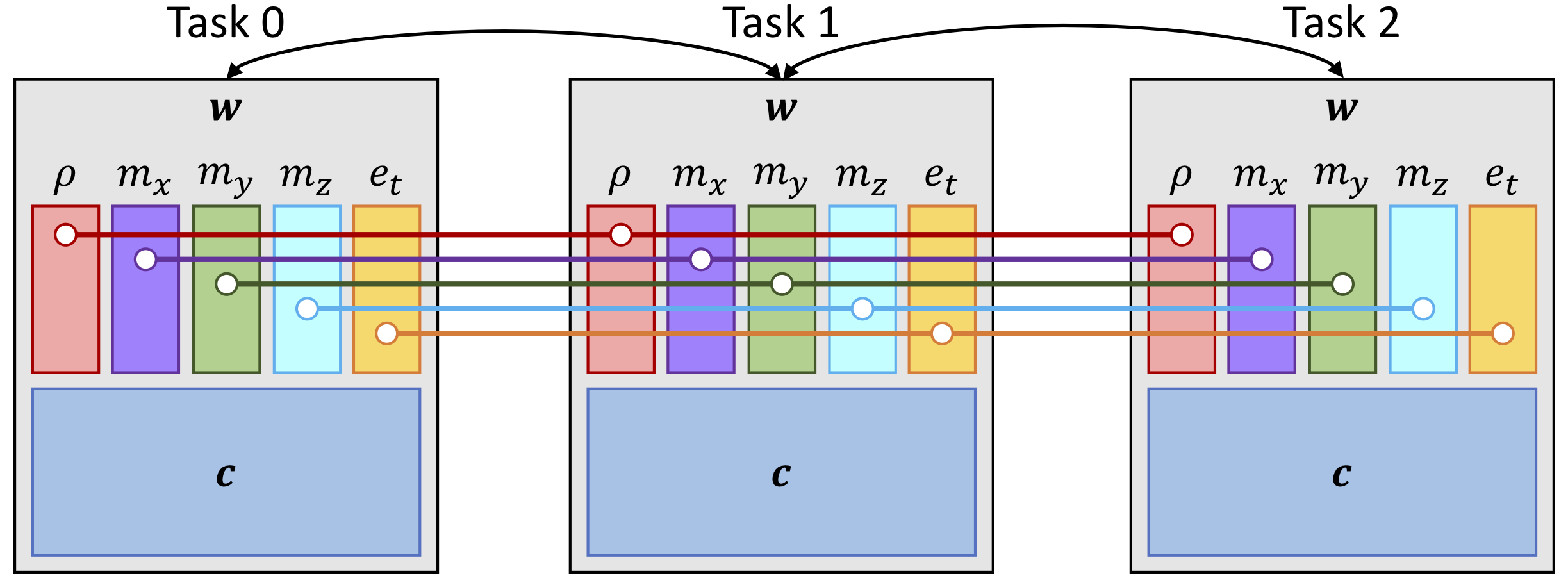}
    \caption{Example illustration of the data partitioning many-vector use case for a reactive flow simulation. The full state, $w$, is composed of subvectors for the fluid variables and chemical species. The fluid quantities (density, $\rho$, momentum in the $x$, $y$, and $z$ directions, $m_*$, and total energy, $e_t$) are stored as parallel vectors each connected by their own communicator. The chemical species, $c$, for local reactions are stored in separate task-specific vectors, e.g., one or more serial, OpenMP, or GPU vectors per task.}
    \label{fig:manyvector1}
\end{figure}

For example, in the data partitioning case on a hybrid CPU+GPU system, a user would create subvectors using existing N\_Vector constructors, some of which may be CPU-specific and others GPU-specific. In this scenario, it is assumed that all MPI-aware subvectors have been constructed using the same MPI intercommunicator.  We note that some of these vectors may represent ``independent'' portions of the data (e.g., for processes that are not coupled between nodes); these can be constructed out of MPI-unaware N\_Vectors, thereby allowing corresponding subvector operations to avoid communication altogether. Once all subvectors have been constructed, the user then combines them together into a single many-vector by passing the subvectors to the many-vector constructor. When performing problem-specific routines (ODE RHS, DAE/nonlinear residual evaluation), the user may then access specific subvectors to retrieve and fill data, potentially copying data between accelerator and host memory as needed. An illustration of this use case for reactive flow is given in Figure \ref{fig:manyvector1}.

The multiphysics use case corresponds to users who wish to combine separate simulations together, e.g., for fluid-structure interaction or atmosphere-ocean coupling in climate simulations, where it is assumed that MPI-aware subvectors have different MPI communicators.  Here, we expect that the user will (a) create the MPI intercommunicator that connects the distinct intracommunicators together, and (b) handle communication operations between intracommunicators to couple the physical models together in their problem-defining nonlinear residual or ODE right-hand side routines.  As in the data partitioning case, the user creates each of the subvectors and the intercommunicator that couples these together. The unifying many-vector is then constructed by providing the subvectors and intercommunicator to the many-vector constructor.

We further note that combinations of these two use cases are also possible via the many-vector interface.  Additionally, a many-vector may be constructed using many-vectors as subvectors, thereby allowing a hierarchy of communication patterns within a simulation.

To achieve the above objectives, the ``MPIManyVector'' implementation \editmade{of an N\_Vector} consists of a relatively lean data structure comprised primarily of an array of subvectors, the global vector length (the combined length of all subvectors for use in RMS-like norms), and an MPI communicator for global reduction operations.
In support of these many-vector modules a new \textit{required} operation, \texttt{N\_VGetLength}, and a new \textit{optional} operation, \texttt{N\_VGetCommunicator}, have been added to the set of vector operations. Additionally, to minimize parallel communication with the MPIManyVector, we have expanded the set of \textit{optional} vector operations to include local reduction kernels so that norm-like calculations may be performed with only a single \texttt{MPI\_Allreduce} call (instead of one for each subvector).
These methods compute only the portions of a reduction operation that are local to a given subvector and MPI rank, and their results are combined by the MPIManyVector into the overall result.  As these local operations are optional, if they are not supplied by a user-provided subvector, then the MPIManyVector implementation will instead compute the final result using the original (required) vector operations (with the corresponding increase in MPI reductions).

For example consider the weighted root-mean-square (WRMS) norm of a vector $x$ with weight vector $w$, both of length $N$, given by
\begin{equation*}
\|x\|_{wrms} = \sqrt{\frac{1}{N}\sum_{i=0}^{N-1}(w_i x_i)^2}.
\end{equation*}
\editmade{In the many-vector case, $x$ and $w$ above are replaced with $X$ and $W$ comprised of $s$ subvectors of length $N_j$ with the total length $N_s = \sum_{j=0}^{s-1} N_j$. The WRMS norm is now given as}
\begin{equation*}
    \|X\|_{wrms} = \sqrt{\frac{1}{N_s}\sum_{j=0}^{s-1}\sum_{i=0}^{N_j-1}\big(W_{j,i} X_{j,i}\big)^2}.
\end{equation*}
The MPIManyVector algorithm for computing the WRMS norm is given in Algorithm~\ref{alg:WRMS}. Here we note the use of 1 versus $s + 1$ global reductions (\textit{each} \texttt{WRMS\_Norm} call requires an \texttt{MPI\_Allreduce}). The full set of these optional local reduction operations can be found in \cite[Section 6.1.4]{cvodeDocumentation}.

\begin{algorithm}[htb!]
\DontPrintSemicolon
\SetAlgoLined
 local\_sum = 0\;
 \ForEach{subvector pair $X_j$, $W_j$}{
  \uIf{local squared sum method exists}{
    local\_sum += Local\_Squared\_Sum($X_j$, $W_j$)\;
  }
  \Else{
    contrib = WRMS\_Norm($X_j$, $W_j$)\;
    \If{subvector root task}{
      subvec\_length = Get\_Length($X_j$)\;
      local\_sum += contrib * contrib * subvec\_length\;
    }
  }
 }
 MPI\_Allreduce(\&local\_sum, \&global\_sum, 1, MPI\_DOUBLE, MPI\_SUM, inter\_comm)\;
 \Return sqrt(global\_sum / global\_vec\_length)\;
 \caption{MPIManyVector weighted root-mean-square (WRMS) norm implementation}
 \label{alg:WRMS}
\end{algorithm}

\subsection{MPIPlusX}
\label{subsec:MPIPlusX}

An added benefit of the MPIManyVector implementation is the ability for users to easily create hybrid MPI+X vectors where ``X'' is a type of on-node parallelism. Given an MPI-unaware on-node vector, either one provided with SUNDIALS (e.g., OpenMP, CUDA, RAJA, etc.) or one defined by the user (e.g., OpenCL, OpenACC, etc.), an MPI+X version of the vector can be simply created by supplying an MPI communicator and the on-node N\_Vector to the MPIManyVector constructor. To streamline this particularly beneficial use case we have created an
``MPIPlusX'' vector, which is a thin wrapper of the MPIManyVector to simplify construction and provide some helpful utility functions for working with the wrapped on-node vector. \editmade{See \cite{balos2020enabling} for a performance comparison of various GPU on-node vectors beneath the MPIPlusX vector.}

\section{Linear Solvers}
\label{sec:linear}

\editmade{The new SUNLinearSolver class serves to streamline the use of problem-specific and third-party linear solvers while maintaining support for integrator-specific linear solver optimizations within SUNDIALS.  To implement a custom SUNLinearSolver, a \textit{minimum} set of required functions must be implemented. Additionally, there are a} variety of optional routines that, \editmade{if provided,} can be leveraged by the SUNDIALS integrators to improve efficiency. \editmade{There are two} required SUNLinearSolver \editmade{methods}: \texttt{SUNLinSolGetType} and \texttt{SUNLinSolSolve} \editmade{(see Table \ref{t:ls_required_ops}).}
\begin{table}[htb!]
\caption{\editmade{Required SUNLinearSolver operations}}
\label{t:ls_required_ops}
\lstset{linewidth=0.65\textwidth}
\small
\begin{tabular}{p{0.65\textwidth}p{0.25\textwidth}}
\toprule
Function & Description \\
\midrule
\begin{codetable}
SUNLinearSolver_Type SUNLinSolGetType(SUNLinearSolver LS)
\end{codetable}
& Returns the solver type \\
\midrule
\begin{codetable}
int SUNLinSolSolve(SUNLinearSolver LS, SUNMatrix A,
                   N_Vector x, N_Vector b, realtype tol)
\end{codetable}
& Solves the linear system $Ax=b$\\
\bottomrule
\end{tabular}
\end{table}
The first of these allows the solver to self-identify its ``type'' as being either: (a) matrix-based and direct, (b) matrix-based and iterative, or (c) matrix-free and iterative.  This identification allows the SUNDIALS packages to exploit certain \editmade{solver} properties for efficiency.  For example, \editmade{with matrix-based linear solvers} the SUNDIALS integrators \editmade{will} infrequently \editmade{update} Jacobian matrices and, for direct linear solvers, matrix factorizations, to amortize the high cost of these operations across multiple time steps or nonlinear solver iterations.  Similarly, \editmade{for iterative linear solvers} all of the SUNDIALS packages \editmade{select} tolerances that are ``just tight enough'' to aid an outer Newton iteration without performing unnecessary linear iterations.  The purpose of the second required routine, \texttt{SUNLinSolSolve}, is self-explanatory -- this performs the solve for a given linear system $Ax=b$.

To support \editmade{integrator-specific optimizations,} the SUNLinearSolver class defines a number of additional ``set'' routines that, if implemented, can be called to provide more fine-grained control over the SUNLinearSolver object.  In addition, a set of ``get'' routines for retrieving artifacts (statistics, residual vectors, etc.) from the linear solver can also be called.  While the SUNDIALS documentation provides an exhaustive list of these set and get routines \cite[Section 8.1]{cvodeDocumentation}, we discuss a few of these \editmade{optional operations (see Table \ref{t:ls_optional_ops})} here to highlight \editmade{their benefits in} the SUNLinearSolver class.
\begin{table}[htb!]
\caption{\editmade{Subset of optional SUNLinearSolver operations}}
\label{t:ls_optional_ops}
\lstset{linewidth=0.65\textwidth}
\small
\begin{tabular}{p{0.65\textwidth}p{0.25\textwidth}}
\toprule
Function & Description \\
\midrule
\begin{codetable}
int SUNLinSolSetup(SUNLinearSolver LS, SUNMatrix A)
\end{codetable}
& Performs infrequent solver setup actions \\
\midrule
\begin{codetable}
int SUNLinSolSetATimes(SUNLinearSolver LS, void* A_data,
                      ATimesFn ATimes)
\end{codetable}
& Provides the solver with a routine to compute the product $Ax \rightarrow z$ \\
\midrule
\begin{codetable}
int SUNLinSolSetScalingVectors(SUNLinearSolver LS,
                              N_Vector s1, N_Vector s2)
\end{codetable}
& Provides the solver with linear system scaling vectors \\
\bottomrule
\end{tabular}
\end{table}

\editmade{The amortization of potentially high cost matrix factorizations or preconditioner setups is achieved} achieved by calling the optional \texttt{SUNLinSolSetup} function with an updated system matrix $A$ (for matrix-based linear solvers).  The frequency of these calls depends on the type of integrator and/or nonlinear solver that uses the linear solver, and that frequency may, in many cases, be modified by the user.  However, if this setup routine is not implemented by the SUNLinearSolver, then these costs must instead be incurred at every call to \texttt{SUNLinSolSolve}, which happens much more frequently. \editmade{To enable this abstraction for matrix-based linear solvers we have created a new SUNMatrix class. This matrix class provides continued support for our pre-existing dense, banded, and sparse-direct linear solvers, and it streamlines interfacing with external linear solver packages.}
\begin{table}[bht!]
\caption{\editmade{Subset of SUNMatrix operations: (R) are required, (O) are optional}}
\label{t:matrix_ops}
\lstset{linewidth=0.5\textwidth}
\small
\begin{tabular}{p{0.5\textwidth}p{0.4\textwidth}}
\toprule
Function & Description \\
\midrule
\begin{codetable}
SUNMatrix SUNMatClone(SUNMatrix A)
\end{codetable}
& (R) Allocates a new matrix with the same size and structure as $A$ \\
\midrule
\begin{codetable}
void SUNMatDestroy(SUNMatrix A)
\end{codetable}
& (R) Frees all memory associated with $A$ \\
\midrule
\begin{codetable}
SUNMatrix_ID SUNMatGetID(SUNMatrix A)
\end{codetable}
& (O) Identifier used by SUNLinearSolvers to check for compatibility \\
\midrule
\begin{codetable}
int SUNMatCopy(SUNMatrix A, SUNMatrix B)
\end{codetable}
& (O) Copies all values from $A$ into $B$ \\
\midrule
\begin{codetable}
int SUNMatScaleAddI(realtype c, SUNMatrix A)
\end{codetable}
& (O) Performs the combination $A\gets cA+I$ \\
\bottomrule
\end{tabular}
\end{table}
\editmade{Here,} each matrix-based SUNLinearSolver implementation has a compatible SUNMatrix implementation, that at a minimum must provide routines to clone or destroy existing SUNMatrix objects, although some SUNDIALS packages require additional matrix operations for compatibility \editmade{(see Table \ref{t:matrix_ops}).}  The set of SUNDIALS-provided SUNMatrix implementations, as well as their compatibility with provided N\_Vector and SUNLinearSolver implementations, is discussed in detail in the SUNDIALS documentation \cite[Section 7.1]{cvodeDocumentation}.

For matrix-free iterative linear solvers, the SUNLinearSolver object requires knowledge of the linear system that must be solved.  As such, linear solvers having this type must implement the \texttt{SUNLinSolSetATimes} routine, that is called by SUNDIALS packages to provide a function and corresponding data structure for computing the matrix-vector product, $Ax \to z$.

Lastly, for \editmade{both matrix-based and matrix-free iterative} solvers, the SUNLinearSolver class defines a set of routines to support both scaling and preconditioning, to balance error between solution components and to accelerate convergence of the linear solver.  To this end, instead of solving a linear system $Ax = b$ directly, iterative SUNLinearSolver implementations may consider the transformed system $\tilde{A} \tilde{x} = \tilde{b}$, where
\begin{align}
  \label{eq:transformed_linear_system_components}
  \tilde{A} = S_1 P_1^{-1} A P_2^{-1} S_2^{-1},\qquad
  \tilde{b} = S_1 P_1^{-1} b,\qquad
  \tilde{x} = S_2 P_2 x,
\end{align}
and where $P_1$ is the left preconditioner, $P_2$ is the right preconditioner, $S_1$ is a diagonal matrix of scale factors for $P_1^{-1} b$, and $S_2$ is a diagonal matrix of scale factors for $P_2 x$.  The SUNDIALS integrators provide these scaling matrices \editmade{such} that $S_1 P_1^{-1} b$ and $S_2 P_2 x$ have dimensionless components.
Furthermore, SUNDIALS packages request that iterative linear solvers stop based on the 2-norm of the scaled preconditioned residual meeting a prescribed tolerance:
\begin{equation*}
  \left\| \tilde{b} - \tilde{A} \tilde{x} \right\|_2  <  \text{tol}.
\end{equation*}
These preconditioning operators and scaling matrices are provided to iterative SUNLinearSolver objects through the optional \texttt{SetPreconditioner} and \texttt{SetScalingVectors} routines.  However, if an iterative linear solver implementation does not support scaling, the SUNDIALS packages will attempt to accommodate for the lack of scaling by instead requesting that the iterative linear solver stop based on the \editmade{criterion}
\begin{equation}
  \label{eq:noscale_tol}
  \left\| P_1^{-1} b - P_1^{-1} A x \right\|_2  <  \text{tol}\, \big/\, \overline{s}_1,
\end{equation}
where $\overline{s}_1$ is the RMS norm of the diagonal of $S_1$. \editmade{This tolerance adjustment} is not a perfect replacement for both solution and residual scaling, in that it only accounts for the magnitude of the overall linear residual and does not balance error between specific residual or solution components.

With this SUNLinearSolver class in place, our redesign operated on two levels. On the SUNDIALS package side, we converted each linear solver interface to utilize generic linear solvers \editmade{and leverage the} three aforementioned types. On the solver side, we converted each existing linear solver in SUNDIALS to an implementation of the SUNLinearSolver \editmade{class.} We note that through the definition and use of these abstract classes alone, we were able to remove nine duplicate linear solver interfaces and three utility files in \textit{each} of the SUNDIALS \editmade{packages,} while simultaneously expanding the set of linear solvers available for each package and maintaining integrator-specific efficiency enhancements. \editmade{The current set of SUNDIALS-provided SUNLinearSolvers includes direct solvers for dense (native, LAPACK \cite{lapack99}, and MAGMA \cite{tdb10}), banded (native and LAPACK), and sparse (KLU \cite{KLU_site,davis2010KLU}, SuperLU\_MT \cite{demmel1999Asynchronous}, SuperLU\_Dist \cite{li2003SuperLU_Dist}, and cuSOLVER \cite{cusolver2019manual}) matrix-based linear systems, as well as an ensemble of matrix-free scaled-preconditioned iterative linear solvers (GMRES, FGMRES, TFQMR, BiCG-Stab, and CG).}

\subsection{\editmade{SUNLinearSolver Demonstration Problem}}
\label{subsec:SUNLinearSolverDemo}

\editmade{To demonstrate the ability for users to supply a custom linear solver for their applications, we consider} the two-dimensional heat equation
\begin{equation*}
    \frac{\partial}{\partial t}u(t,x,y) = \nabla \cdot (D \nabla u(t,x,y)) + b(t,x,y)
\end{equation*}
where $u$ is the temperature, $D$ is a diagonal matrix containing the constants \editmade{$k_x = k_x = 1.0$} for the diffusivity in the x- and y-directions respectively, and the external forcing term $b(t,x,y)$ is given by
\begin{equation*}
\begin{split}
    b(t,x,y) = &-2 \pi \sin^2(\pi x) \sin^2(\pi y) \sin(\pi t) \cos(\pi t) \\
    &- k_x 2 \pi^2 (\cos^2(\pi x) - \sin^2(\pi x)) \sin^2(\pi y) \cos^2(\pi t) \\
    &- k_y 2 \pi^2 (\cos^2(\pi y) - \sin^2(\pi y)) \sin^2(\pi x) \cos^2(\pi t).
\end{split}
\end{equation*}
The problem is solved on the unit square $(x,y) \in [0,1]^2$ for $t \in [0,1]$ with $u(t,x,y) = 0$ on the boundary, and the initial condition is $u(0,x,y) = \sin^2(\pi x) \sin^2(\pi y)$. With this configuration, the problem has the analytical solution $u(t,x,y) = \sin^2(\pi x) \sin^2(\pi y) \cos^2(\pi t)$.

\editmade{Spatial} derivatives are discretized using second order centered finite differences with data distributed over an $n_x \times n_y$ uniform spatial grid \editmade{decomposed across a two-dimensional Cartesian communicator with $n_{px} \times n_{py}$ MPI tasks. We integrate this problem in time using both CVODE and ARKODE; the} ARKODE version of the \editmade{problem uses the} default third order diagonally implicit Runge-Kutta method in ARKODE, while the CVODE version uses adaptive order BDF methods.
\editmade{For both examples, the implicit systems(s) in each time step are solved using SUNDIALS' default inexact Newton method with custom} SUNMatrix and SUNLinearSolver implementations that wrap a structured grid matrix and iterative linear solvers from \editmade{\textit{hypre}, CG and GMRES.}
The linear \editmade{solvers} are preconditioned with \textit{hypre}'s PFMG parallel semicoarsening multigrid solver \cite{ashby1996parallel, falgout2000multigrid}.
\editmade{The source files for the CVODE and ARKODE versions of this \editmade{example} problem are
\href{https://github.com/LLNL/sundials/blob/master/examples/cvode/CXX_parhyp/cv_heat2D_hypre_ls.cpp}{\wraptt{cv_heat2D_hypre_ls.cpp}}
and
\href{https://github.com/LLNL/sundials/blob/master/examples/arkode/CXX_parhyp/ark_heat2D_hypre_ls.cpp}{\wraptt{ark_heat2D_hypre_ls.cpp}}
respectively.}

\editmade{Here, the member data for the custom SUNMatrix object consists of the \texttt{Hypre5ptMatrixContent} structure that} holds the \textit{hypre} structured grid, stencil, and matrix objects as well as workspace arrays and other information for updating the matrix entries. \editmade{A similar structure (\texttt{HypreLSContent}) defines the member data for the SUNLinearSolver.
Since} the linear solver is an iterative method that uses a matrix object, the \editmade{custom} SUNMatrix implementation only needs to define the \texttt{GetID}, \texttt{Clone}, \texttt{Destroy}, \texttt{Copy}, and \texttt{ScaleAddI} methods \editmade{from Table \ref{t:matrix_ops}}.
The \editmade{custom} linear solver defines the required \texttt{GetType} and \texttt{Solve} methods \editmade{as well as} the optional \texttt{Setup} method to enable reusing the preconditioner \editmade{setup.}
Additionally, the \editmade{custom} linear solver implementation provides the optional \texttt{NumIters} and \texttt{Free} methods for retrieving the number of iterations from a solve and deallocating the SUNLinearSolver and its content.

\editmade{Since this custom} solver does not provide a function to set scaling vectors, the \editmade{integrators adjust} the linear solve tolerance \editmade{to compensate, as shown in \eqref{eq:noscale_tol}. The} main function \editmade{for each example} creates SUNMatrix and SUNLinearSolver instances using the \editmade{custom} constructors, \texttt{Hypre5ptMatrix} and \texttt{HypreLS}, respectively. \editmade{Both of these} constructors allocate the base object structure with the corresponding empty constructor (as described in Section \ref{sec:structure}), set the function pointers in the \editmade{VMT}, allocate the object content and attach the content structure, and\editmade{,} finally\editmade{,} create/initialize the content data. Once the matrix and linear solver objects are created, the \editmade{example programs attach} the objects to the integrators with the ARKODE and CVODE \texttt{SetLinearSolver} functions.

\begin{table}
    \caption{\editmade{Integrator statistics from the SUNLinearSolver demonstration problem in Section \ref{subsec:SUNLinearSolverDemo}. The MF and JV columns correspond to SUNDIALS solvers with a matrix-free or a user-defined Jacobian-vector product, respectively.  The statistics given are the number of successful time steps (Steps), failed time steps (Step fails), total (integrator + linear solver) ODE right-hand side evaluations (RHS evals), nonlinear iterations (NLS iters), and linear iterations (LS iters). The max error at the final time (Error) and the total average run time across 100 runs (Run time) are also given.}}

    \label{t:2dheat}
    \begin{tabular}{lccc|ccc||c|c}
        \toprule
        & \multicolumn{6}{c||}{SUNDIALS v5.5} & \multicolumn{2}{c}{SUNDIALS v2.7} \\
        \midrule
        & \multicolumn{3}{c|}{CVODE} & \multicolumn{3}{c||}{ARKODE} & CVODE$^a$ & ARKODE$^b$ \\
        \midrule
        \multirow{2}{*}{CG} & \multirow{2}{*}{\textit{hypre}} & SUN & SUN & \multirow{2}{*}{\textit{hypre}} & SUN & SUN & SUN & SUN \\
                             &                                 & MF  & JV  &                                 & MF  & JV  & MF  & MF  \\
        \midrule                
        Steps                   & 84        & 84         & 84         & 148        & 144         & 144         & --      & 144    \\
        Step fails              & 4         & 4          & 4          & 2          & 1           & 1           & --      & 1      \\
        RHS evals               & 109       & 609        & 109        & 1,053      & 5,253       & 1,018       & --      & 5,250  \\
        NLS iters               & 106       & 106        & 106        & 451        & 435         & 435         & --      & 435    \\
        LS iters                & 449       & 500        & 500        & 4,276      & 4,235       & 4,236       & --      & 4,231  \\
        Error $(\times10^{-7})$ & 3.2       & 3.2        & 3.2        & 4.0        & 3.5         & 3.5         & --      & 3.7    \\
        Run time (sec)          & 8.8       & 13.0       & 9.3        & 67.3       & 102.0       & 70.8        & --      & 104.2  \\
        \midrule
        \multirow{2}{*}{GMRES} & \multirow{2}{*}{\textit{hypre}} & SUN & SUN & \multirow{2}{*}{\textit{hypre}} & SUN & SUN & SUN & SUN \\
                               &                                 & MF  & JV  &                                 & MF  & JV  & MF  & MF  \\
        \midrule                
        Steps                   & 84        & 80         & 80         & 147        & 144         & 144         & 80      & 144    \\
        Step fails              & 4         & 4          & 4          & 2          & 1           & 1           & 4       & 1      \\
        RHS evals               & 109       & 531        & 104        & 1,042      & 5,193       & 1,018       & 532     & 5,196  \\
        NLS iters               & 106       & 101        & 101        & 446        & 435         & 435         & 101     & 435    \\
        LS iters                & 394       & 427        & 427        & 4,014      & 4,175       & 4,174       & 427     & 4,177  \\
        Error $(\times10^{-7})$ & 3.2       & 11.1       & 11.1       & 3.8        & 3.6         & 3.6         & 11.1    & 3.7    \\
        Run time (sec)          & 8.2       & 12.7       & 9.4        & 67.5       & 109.7       & 78.9        & 12.3    & 108.6  \\
        \bottomrule
        \multicolumn{9}{l}{
        \footnotesize{$^a$ An interface to CG from CVODE is not available in SUNDIALS v2.7.0}}\\
        \multicolumn{9}{l}{
        \footnotesize{$^b$ A patch is required to correct the norm used to test for CG convergence in SUNDIALS v2.7.0.}}
    \end{tabular}
\end{table}

\editmade{The updated linear solver interfaces make it easy to quickly compare multiple solvers and integrators on a problem. Results comparing the CG and GMRES linear solvers from SUNDIALS (matrix-free or with a user-supplied Jacobian-vector product function) and \textit{hypre} (matrix-based), each preconditioned with PFMG, underneath a linear multistep integrator (CVODE) and a multistage integrator (ARKODE) are given in Table \ref{t:2dheat}. See} \editmade{the files} \href{https://github.com/LLNL/sundials/blob/master/examples/arkode/CXX_parhyp/ark_heat2D_hypre_pfmg.cpp}{\texttt{ark\_heat2D\_hypre\_pfmg.cpp}} and \href{https://github.com/LLNL/sundials/blob/master/examples/cvode/CXX_parhyp/cv_heat2D_hypre_pfmg.cpp}{\texttt{cv\_heat2D\_hypre\_pfmg.cpp}} for the versions of this example \editmade{that use} the SUNDIALS linear solvers. Additionally, Table \ref{t:2dheat} includes results from SUNDIALS v2.7.0\footnote{The implementation of CG in SUNDIALS v2.7.0 incorrectly used a \editmade{WRMS} norm rather than a weighted 2-norm for checking convergence. As such the results presented are with a patched version of v2.7.0 updated to use the correct norm.} before the introduction of the new \editmade{SUNLinearSolver class} to illustrate that solver performance is unaffected \editmade{by the new infrastructure (i.e., run times are within the observed $\sim\!2\%$ variability).} As only ARKODE had an interface to the SUNDIALS CG solver before the introduction of the linear solver class, results with CVODE from SUNDIALS v2.7.0 only utilize the SUNDIALS matrix-free GMRES implementation.

\editmade{The} tests use a 4,096 $\times$ 4,096 spatial grid distributed across 256 MPI tasks (16 tasks each in the x- and y-directions) and are run on 8 nodes of the Quartz cluster at Lawrence Livermore National Laboratory. Each compute node is composed of two Intel Xeon E5-2695 v4 chips, each with 18 cores. MPI tasks are distributed equally across the compute nodes so \editmade{that} 32 of the 36 cores available on a node are utilized. All integrator and solver options are set to the default \editmade{values and} have not been optimized for this \editmade{example demonstrating} the re-usability of class implementations.

In this example, as seen from Table~\ref{t:2dheat}, the combination of CVODE and a matrix-based linear solver from \textit{hypre} produces the most efficient results for the desired accuracy (the integrator relative and absolute tolerances are $10^{-5}$ and $10^{-10}$ respectively). CVODE is well suited to this problem as it only requires one nonlinear solve per step, regardless of the method order, and adapts the order in addition to the time step size to maximize efficiency. ARKODE also adapts the step size, but uses a fixed order \editmade{method which} requires \editmade{multiple} implicit solves per step (for a problem better suited to the multistage IMEX methods in ARKODE see the 1D advection-reaction problem \editmade{in Section \ref{subsec:SUNNonlinearSolverDemo}}).  With the default parameters, the matrix-based \textit{hypre} solvers lead to faster run times than the SUNDIALS solvers. The increased cost in the matrix-free case is a result of the additional right-hand side evaluations (and the corresponding MPI communication) required to approximate the Jacobian-vector product. Supplying a user-defined Jacobian-vector product function eliminates this cost. However, the SUNDIALS solvers still require approximately 0.5 more linear iterations per nonlinear iteration. The lower average number of iterations with \textit{hypre} solvers is, in part, due to the tolerance adjustment \editmade{\eqref{eq:noscale_tol}} for linear solvers that do not support scaling. The performance of the SUNDIALS linear solvers can be improved for this problem by relaxing the linear solve \editmade{tolerance.}

\editmade{These results show the flexibility of the new SUNLinearSolver class to enable comparisons of various integrator and solver options to determine the best combination of integrators, solvers, and parameters for a given application. Prior to the introduction of the new matrix and linear solver classes, such a comparison across different integrators would require the user to implement unique interfaces to \textit{hypre} solvers for both CVODE and ARKODE. With the new SUNDIALS infrastructure such interfaces can easily be reused between the integrators. Moreover, the user's calling program to setup and evolve the problem in each of these cases is essentially identical. To switch from a native SUNDIALS linear solver to using a \textit{hypre} solver, the user only needs to swap the SUNDIALS constructor call with a call to their matrix and linear solver constructors and update the \texttt{SetLinearSolver} call to also attach a matrix object. To enable the user-defined Jacobian-vector product, only one additional function call is necessary to attach the user's function. All other linear solver related functions, e.g., adjusting the preconditioner/linear system update frequency or linear solve tolerance, are independent of the choice of SUNLinearSolver object.}

\section{Nonlinear Solvers}
\label{sec:nonlinear}

\editmade{The new SUNNonlinearSolver class provides an interface for supplying SUNDIALS integrators with problem-specific and third-party nonlinear solvers. Like the other SUNDIALS classes, the SUNNonlinearSolver base class defines a set of required methods for solving nonlinear systems and an additional set of optional methods that can be leveraged to improve solver efficiency, set solver options, or retrieve solver information. Complete descriptions of all SUNNonlinearSolver methods can be found in the SUNDIALS documentation \cite[Section 9.1]{cvodeDocumentation}.}
\begin{table}[htb]
\caption{\editmade{Required SUNNonlinearSolver operations}}
\label{t:nls_required_ops}
\lstset{linewidth=0.60\textwidth}
\small
\begin{tabular}{p{0.60\textwidth}p{0.30\textwidth}}
\toprule
Function & Description \\
\midrule
\begin{codetable}
SUNNonlinearSolver_Type
  SUNNonlinSolGetType(SUNNonlinearSolver NLS)
\end{codetable}
& Returns the solver type \\
\midrule
\begin{codetable}
int SUNNonlinSolSetSysFn(SUNNonlinearSolver NLS,
                         SUNNonlinSolSysFn SysFn)
\end{codetable}
& Set a function pointer to the nonlinear system function\\
\midrule
\begin{codetable}
int SUNNonlinSolSolve(SUNNonlinearSolver NLS,
                      N_Vector y0, N_Vector ycor,
                      N_Vector w, realtype tol,
                      booleantype callLSetup, void *mem)
\end{codetable}
& Solves the nonlinear system \\
\bottomrule
\end{tabular}
\end{table}

\editmade{The three required nonlinear solver methods are listed in Table \ref{t:nls_required_ops}. These include the \texttt{GetType} method for querying a solver object about the nonlinear system formulation it expects: root-finding, $F(u) = 0$, or fixed-point, $G(u) = 0$. Based on this information, an integrator will use the \texttt{SetSysFn} method to supply the solver with a function to evaluate the appropriate nonlinear system. As nonlinear systems in SUNDIALS integrators are formulated in terms of a correction to the predicted solution, the \texttt{Solve} method is passed the initial guess for the state vector, \texttt{y0}, as well as the correction vector, \texttt{ycor}, which contains the initial correction on input and the final correction on output. The solution is expected to satisfy the tolerance, \texttt{tol}, in the WRMS norm with weight vector, \texttt{w}. As discussed in Section~\ref{sec:linear}, nonlinear solvers like Newton's method may reuse linear system information between solves to reduce the cost of setting up the linear solver. The input flag, \texttt{callLSetup}, signals whether the linear system matrix and/or preconditioner should be updated.}

\editmade{Rather than relying directly on the SUNLinearSolver class, the optional \texttt{SetLSetupFn} and \texttt{SetLSolveFn} methods listed in Table \ref{t:nls_optional_ops} allow the integrators to provide the nonlinear solver with functions handling integrator-specific actions for setting up and solving linear systems. These functions in turn utilize SUNLinearSolver objects. For nonlinear solvers that do not require a linear solver, or the solve is done within the nonlinear solver implementation, these optional methods do not need to be provided.
Additionally, the optional \texttt{SetConvTestFn} method listed in Table \ref{t:nls_optional_ops} allows an integrator-specific convergence test function to be provided.} The default stopping test for nonlinear iterations is related to the subsequent local error test for a time step, with the goal of keeping the nonlinear iteration errors from interfering with local temporal error control. In each of the integrators, this test is based on an estimate of the solver convergence rate, and incorporates several heuristics for estimating the rate of convergence and detecting solver divergence. As the default convergence test may not be optimal for all nonlinear solvers or problems, this optional \editmade{method} also allows a user to attach a custom convergence test that is better-suited to the solver or application. Alternately, if the nonlinear solver includes its own internal convergence test, this optional function need not be provided.
\begin{table}[bht]
\caption{\editmade{Subset of optional SUNNonlinearSolver operations}}
\label{t:nls_optional_ops}
\lstset{linewidth=0.65\textwidth}
\small
\begin{tabular}{p{0.65\textwidth}p{0.20\textwidth}}
\toprule
Function & Description \\
\midrule
\begin{codetable}
int SUNNonlinSolSetLSetupFn(SUNNonlinearSolver NLS,
                            SUNNonlinSolLSetupFn SetupFn);
\end{codetable}
& Set linear solver setup function \\
\midrule
\begin{codetable}
int SUNNonlinSolSetLSolveFn(SUNNonlinearSolver NLS,
                            SUNNonlinSolLSolveFn SolveFn);
\end{codetable}
& Set linear solver solve function \\
\midrule
\begin{codetable}
int SUNNonlinSolSetConvTestFn(SUNNonlinearSolver NLS,
                              SUNNonlinSolConvTestFn CTestFn,
                              void* ctest_data);
\end{codetable}
& Set convergence test function \\
\bottomrule
\end{tabular}
\end{table}

With the new SUNNonlinearSolver class in place, \editmade{the integrators were updated to utilize generic nonlinear solvers, and the existing solver implementations were combined into shared derived classes. This change greatly improves the flexibility of the time integrators, as third party nonlinear solver interfaces (e.g., the SUNNonlinearSolver\_PetscSNES class utilizing PETSc’s SNES solver \cite{petsc-web-page,petsc-user-ref}), or custom nonlinear solvers (as illustrated below), can now be created.}

\subsection{\editmade{SUNNonlinearSolver Demonstration Problem}}
\label{subsec:SUNNonlinearSolverDemo}

\editmade{To demonstrate the flexibility provided by the new nonlinear solver class we consider a one-dimensional advection-reaction problem using a stiff variation of the Brusselator model from chemical kinetics \cite{hairer1996solving} utilizing a custom SUNNonlinearSolver. The system is given by}
\begin{align*}
    \frac{\partial u}{\partial t} &= -c \frac{\partial u}{\partial x} + A - (w+1) u + v u^2 \\
    \frac{\partial v}{\partial t} &= -c \frac{\partial v}{\partial x} + w u - v u^2 \\
    \frac{\partial w}{\partial t} &= -c \frac{\partial w}{\partial x} + \frac{B-w}{\epsilon} - w u
\end{align*}
where $u$, $v$, and $w$ are chemical concentrations, \editmade{$t \in [0, 10]$} is time, \editmade{$x \in [0, b]$} is the spatial variable, $c = 0.01$ is the advection speed, $A = 1$ and $B = 3.5$ are the concentrations of chemical species that remain constant over space and time, and $\epsilon = 5 \times 10^{-6}$ is a parameter that \editmade{determines} the stiffness of the system. The problem \editmade{uses periodic boundary conditions and the} initial condition is
\begin{equation*}
\begin{gathered}
    u(0,x) = A + p(x), \qquad
    v(0,x) = B/A + p(x), \qquad
    w(0,x) = 3.0 + p(x),\\
    p(x) = 0.1 \exp\left(-2\left(2x/b-1\right)^2\right).
\end{gathered}
\end{equation*}
\editmade{Spatial} derivatives are discretized with first order upwind finite differences on a uniform mesh with $n_x$ \editmade{points divided} across $n_{px}$ MPI tasks. \editmade{The state is stored in a task-local serial vector wrapped by the MPIPlusX vector.} The default third order IMEX method from ARKODE is used to evolve the system in time with the advection terms treated explicitly and the reaction terms implicitly.

\editmade{As} the reactions are purely local \editmade{in space, the implicit system that must be solved at each stage can be decomposed into independent nonlinear systems at each grid point. To exploit this locality, the user code implements a \textit{custom problem-specific} SUNNonlinearSolver to perform MPI task-local solves with the local serial vectors. This eliminates} nearly all parallel communication within the implicit solves, i.e., global reductions for norms and orthogonalization in GMRES. The only global communication necessary is at the end of the solve phase to check if all the task-local nonlinear solves were successful. \editmade{The source file for this example is \href{https://github.com/LLNL/sundials/blob/master/examples/arkode/C_parallel/ark_brusselator1D_task_local_nls.c}{\wraptt{ark_brusselator1D_task_local_nls.c}.}}

\editmade{Here, the custom nonlinear solver member data is defined as the \texttt{TaskLocalNewton\_Content} structure containing an instance of the native SUNDIALS Newton solver and other data for solving the local nonlinear problem. The custom solver defines the required \texttt{GetType}, \texttt{SetSysFn}, and \texttt{Solve} methods as well as the optional methods: \texttt{SetConvTestFn} to use the integrator-provided convergence test function, \texttt{Free} to deallocate the solver, and \texttt{GetNumConvFails} to retrieve the number of solver failures. Additionally, the optional \texttt{Initialize} method is used, in this case, to provide the task-local instance of the SUNDIALS Newton solver with user-defined functions to evaluate the local residual (\texttt{TaskLocalNlsResidual}) and solve the local linear systems (\texttt{TaskLocalLSolve}). The main function creates a solver instance using the custom constructor, \texttt{TaskLocalNewton}. This function allocates the base object structure with the empty constructor (as described in Section \ref{sec:structure}), sets the function pointers in the VMT, allocates the object content and attaches the content structure, and, finally, creates/initializes the content data. Once the custom solver object is created, the example program attaches it to the integrator with \texttt{ARKStepSetNonlinearSolver}.}

\editmade{To compare this solution approach with what was possible before the addition of the SUNNonlinearSolver class, the example code provides an option to use the SUNDIALS Newton method to perform a global solve with GMRES as the linear solver and a user-defined, task-local preconditioner. Multiple tests are run with a fixed spatial resolution, $b/n_{x} = 10^{-3}$, and the domain size is increased with the number of MPI tasks; specifically we test with $b \in \{1, 2, 4, 8, 16, 32, 64, 128, 256\}$ and $n_{px} \in \{40, 80, 160, 320, 640, 1280, 2560, 5120, 10240\}$ with 40 MPI tasks per compute node on the Lassen supercomputer at Lawrence Livermore National Laboratory. Note the results below utilize only the CPUs (two IBM Power9 processors with 44 cores total) on Lassen. For results utilizing GPU acceleration and the MPIPlusX vector with various GPU task-local vectors see \cite{balos2020enabling}.}

\begin{figure}[htb!]
    \centering
    \includegraphics[trim={0 0 0 0.90cm},clip,width=0.59\linewidth]{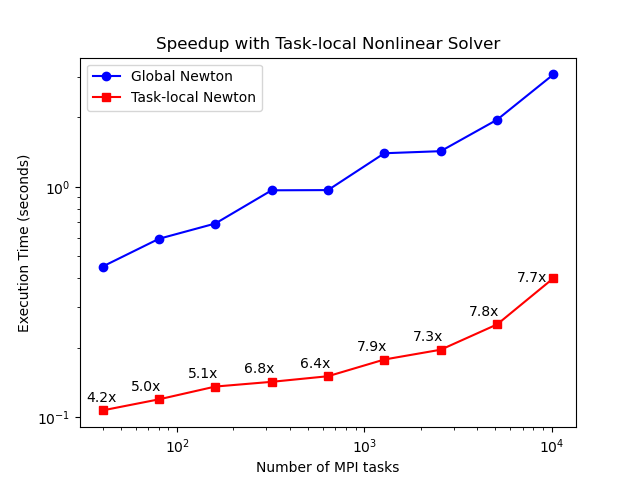}
    \caption{\editmade{Speedup results for various problem sizes for the demonstration problem in Section \ref{subsec:SUNNonlinearSolverDemo} using a custom task-local nonlinear solver (red squares) and a global Newton solve (blue circles). Execution times are the average of 20 runs and each task-local solver data point is annotated with the approximate speedup achieved compared to the global solver.}}
    \label{fig:weakscale-bruss}
\end{figure}

\editmade{Execution times for the global and task-local solvers are given in Figure \ref{fig:weakscale-bruss}. Although the reduction in synchronizations does not significantly impact the problem scaling, the custom solver results in speedups of 4x to almost 8x.} As such, it is clear that the new nonlinear solver class provides important new capabilities that were not possible before the flexibility enhancements and can lead to significantly more efficient solver approaches by exploiting problem structure.

\section{Fortran 2003 Interfaces}
\label{sec:f2003}

Fortran has historically been heavily used in scientific computing and remains widespread today. As such, although SUNDIALS is written \editmade{in C, Fortran interfaces have} been provided for many years. These Fortran 77 standard compliant interfaces have allowed Fortran users to access a \editmade{\textit{subset}} of features from CVODE, IDA, ARKODE, and KINSOL \editmade{and a \textit{limited} number} of the \editmade{vector, matrix, and solver class implementations available to C/C++ users.} However, \editmade{these interfaces} had limited flexibility and sustainability. Notably, the lack of derived types in Fortran 77 was incompatible with the benefits of the SUNDIALS classes.  To circumvent this incompatibility, the \editmade{previous interfaces} accessed SUNDIALS classes using global memory, rendering them as non-threadsafe. As a result, it was difficult for \editmade{Fortran users} to supply application-specific data layouts or custom solvers, and it was impossible to use multiple instances of SUNDIALS packages simultaneously in a multithreaded environment. Furthermore, maintaining these interfaces required significant manual upkeep of code to ``glue'' the interfaces to C; thus, the Fortran 77 interfaces were not sustainable in the long-term.

To address these limitations, we have introduced new Fortran 2003 standard compliant interfaces that provide access to \editmade{\textit{all of the features in all six} of the SUNDIALS packages (access to CVODES and IDAS were previously unsupported from Fortran 77) as well as most of the core class implementations.} These new interfaces are generated automatically using the SWIG-Fortran \editmade{\cite{johnson2019swig}} tool, which only requires minimal additional interface code, much of which is common across SUNDIALS, to create bindings using the Fortran 2003 features for interoperability with ISO C code. When the C API changes in any package or class with an existing Fortran 2003 interface, the SWIG-Fortran tool is run, and the interface ``glue'' code is regenerated before package release. This process fits nicely into a modern software development workflow and \editmade{could be} further automated to run as part of a continuous integration/continuous delivery pipeline. Thus, the Fortran 2003 interfaces have a low-maintenance cost and can quickly be extended to support new features and class implementations as they are \editmade{added, i.e.,} the Fortran 2003 interfaces are sustainable.

Since the Fortran 2003 interfaces are based on features of the Fortran 2003 standard for interopability with ISO C code, they allow Fortran users to continue to write idiomatic Fortran with few caveats, but to then interact with SUNDIALS in a way that closely resembles the C API. In fact, the Fortran 2003 interface API is close enough to the C API that the C API documentation serves dual-purpose as the Fortran 2003 documentation, modulo some information on data types and subtleties attributed to inherent differences of C and Fortran. Function signatures match the C API, with an ``F'' prepended to the function name, e.g. \texttt{FN\_VConst} instead of \texttt{N\_VConst}. The generic SUNDIALS classes, such as the \texttt{N\_Vector}, are interfaced as Fortran derived-types of the same name. This convention means that users can easily provide custom, application-specific implementations of the generic SUNDIALS classes that are written entirely in Fortran \editmade{(as illustrated below)}. Data arrays, e.g., real arrays, are interfaced as Fortran arrays of the corresponding type defined in the \texttt{iso\_c\_binding} intrinsic module. The exact mapping of SUNDIALS C types to Fortran 2003 types, as well as discussion of subtle usage differences between the C and Fortran 2003 APIs, is given in the SUNDIALS documentation \cite[Section 5.1]{cvodeDocumentation}.

\subsection{\editmade{Fortran 2003 Interface Demonstration Problem}}
\label{subsec:F2003Demo}

A second version of the \editmade{demonstration problem from Section \ref{subsec:SUNNonlinearSolverDemo} utilizing a custom nonlinear solver is also implemented in Fortran. The source code for this version of the problem is \href{https://github.com/LLNL/sundials/blob/master/examples/arkode/F2003_parallel/ark_brusselator1D_task_local_nls_f2003.f90}{\wraptt{ark_brusselator1D_task_local_nls_f2003.f90}}. As the Fortran interface closely mirrors the C API, the same steps to create a derived class implementation apply to the Fortran use case.}

In this example, all of the data needed by the \editmade{custom task-local nonlinear solver is contained in a user-defined module, \texttt{nls\_mod}. As such, the data is effectively global (albeit namespace protected). An alternative, thread safe approach would be to define a new type for the solver data and attach it as the class member data. In this case, the module includes an instance of the SUNDIALS Newton method and other data needed for solving the task-local nonlinear systems as well as functions implementing required and optional nonlinear solver methods. An instance of the nonlinear solver object is created using the constructor subroutine \texttt{TaskLocalNewton} which leverages the corresponding base class constructor to create a SUNNonlinearSolver object and \texttt{c\_funloc} to set the function pointers in the VMT. In this case, the object's content field does not need to be set as all the data needed by the solver is accessible through the \texttt{nls\_mod} module. Once the solver and all its required data is created, the nonlinear solver object is attached to the integrator with the \texttt{FARKStepSetNonlinearSolver} function. As expected, the C and Fortran versions of the problem have the same integrator and solver statistics and nearly identical run times.}

\section{Conclusions}
\label{sec:conclusions}

The growing complexity of scientific simulations and recent advances in high performance computing systems have resulted in increased demand for utilizing \editmade{new programming models and more varied algebraic solver options with the SUNDIALS time integrators and nonlinear solvers.} To meet these demands, we have \editmade{created new matrix, linear solver, and nonlinear solver classes following the same object-oriented design previously utilized by the SUNDIALS vector.} These new additions adhere to a number of guiding principles that provide greater flexibility while increasing the sustainability of the SUNDIALS packages, requiring minimal changes to user codes, and resulting in significant performance benefits in many situations.

The new classes allow for easier interoperability with and reuse of external and application-specific data structures and solvers across SUNDIALS packages. In particular, the encapsulation of the nonlinear solvers from the time integrators provides flexibility in nonlinear solution approaches that was not previously available in SUNDIALS. Additionally, we have expanded the SUNDIALS vector class implementations to include a new many-vector capability, allowing for a vector comprised of various subvectors. This many-vector construct provides an easy way to accommodate multiphysics simulations as well as heterogeneous architectures where different vectors operate on data residing in potentially different memory spaces. An MPI+X vector was further added on top of the many-vector to streamline the introduction of new vectors supporting various approaches to on-node parallelism with heterogeneous computing platforms. \editmade{New fused operations better utilize available hardware and enable development of reduced communication algorithms.} Lastly, modern Fortran interfaces allow full access to the flexibility provided with the SUNDIALS classes.

Numerical tests have shown that the new infrastructure retains the existing behavior of the time integrators and does not increase the cost of simulations. Moreover, the results demonstrate the power of being able to easily compare the performance of fundamentally different integrators and solver approaches within the same problem. Significant speedup results were also shown when leveraging the new nonlinear solver class to provide a problem-specific solver that can exploit the structure and locality in the nonlinear system.

With these new capabilities in place, future work includes implementing new vectors \editmade{and matrices} to accommodate more types of on-node parallelism, \editmade{investigating algorithms that leverage fused operations for greater performance, and expanding the set of native solver interfaces provided with SUNDIALS.
Additionally, we plan to leverage the new many-vector implementation within SUNDIALS to exploit the structural features of different applications.}

\begin{acks}
The authors thank Alan Hindmarsh for numerous discussions and insights into details of integrator controls on solvers and design of integrator codes.
\end{acks}

\bibliographystyle{ACM-Reference-Format}
\bibliography{sources}




\end{document}